\documentclass{PoS}
\usepackage{amsmath,amssymb,bm,longtable,mathrsfs,slashed}

\title{\begin{picture}(0,0)(0,0)%
   \put(360,68){\makebox(0,0)[l]{\textnormal{\normalsize OU-HET-1039}}}%
\end{picture}Axial U(1) symmetry and mesonic correlators at high temperature in $N_f=2$ lattice QCD}

\ShortTitle{Axial U(1) symmetry and mesonic correlators at high temperature in $N_f=2$ lattice QCD}

\author{\speaker{Kei Suzuki}$\, ^a$, Sinya Aoki$\, ^b$, Yasumichi Aoki$\, ^{c}$, Guido Cossu$\, ^d$, Hidenori Fukaya$\, ^e$, Shoji Hashimoto$\, ^{f,g}$, Christian Rohrhofer$\, ^e$ \ (JLQCD Collaboration)\\
        \llap{$^a$}Advanced Science Research Center, Japan Atomic Energy Agency (JAEA), Tokai 319-1195, Japan\\
        \llap{$^b$}Center for Gravitational Physics, Yukawa Institute for Theoretical Physics, Kyoto 606-8502, Japan\\
        \llap{$^c$}RIKEN Center for Computational Science, Kobe 650-0047, Japan\\
        \llap{$^d$}School of Physics and Astronomy, The University of Edinburgh, Edinburgh EH9 3JZ, United Kingdom\\
        \llap{$^e$}Department of Physics, Osaka University, Toyonaka 560-0043, Japan\\
        \llap{$^f$}KEK Theory Center, High Energy Accelerator Research Organization (KEK), Tsukuba 305-0801, Japan\\
        \llap{$^g$}School of High Energy Accelerator Science, The Graduate University for Advanced Studies (Sokendai), Tsukuba 305-0801, Japan\\
}

\abstract{%
We investigate the high-temperature phase of QCD using lattice QCD simulations with $N_f = 2$ dynamical M\"obius domain-wall fermions.
On generated configurations, we study the axial $U(1)$ symmetry, overlap-Dirac spectra, screening masses from mesonic correlators, and topological susceptibility. 
We find that some of the observables are quite sensitive to lattice artifacts due to a small violation of the chiral symmetry.
For those observables, we reweight the M\"obius domain-wall fermion determinant by that of the overlap fermion.
We also check the volume dependence of observables.
Our data near the chiral limit indicates a strong suppression of the axial $U(1)$ anomaly at temperatures $\geq$ 220 MeV.
}

\FullConference{37th International Symposium on Lattice Field Theory - Lattice2019\\
		16-22 June 2019\\
		Wuhan, China}

\begin{document}
\vspace{-5pt}
\section{Introduction}\label{sec-1}
\vspace{-5pt}
In the high-temperature region of quantum chromodynamics (QCD), one of open questions is the fate of the $U(1)_A$ symmetry.
In the low-temperature phase, the $U(1)_A$ symmetry is known to be broken by a quantum anomaly which is related to topological excitations of gluon fields, e.g, instantons.
In the high-temperature region with restored chiral symmetry (in other words, above the critical temperature, $T>T_c$), the restoration or violation of the $U(1)_A$ symmetry is still a long-standing problem not only in theoretical approaches \cite{Cohen:1996ng,Aoki:2012yj,Kanazawa:2015xna} but also in lattice QCD simulations at $N_f=2$ \cite{Cossu:2013uua,Chiu:2013wwa,Brandt:2016daq,Tomiya:2016jwr,Ishikawa:2017nwl} and $N_f=2+1$ \cite{Bazavov:2012qja,Buchoff:2013nra,Bhattacharya:2014ara,Dick:2015twa,Mazur:2018pjw,Bazavov:2019www}.

In older studies, lattice simulations reported a sizable $U(1)_A$ symmetry breaking above the critical temperature.
However, many studies applied the staggered-type fermions, where chiral symmetry is explicitly broken, and it was difficult to precisely measure how much of $U(1)_A$ symmetry breaking is due to lattice artifacts.
Recently, chiral fermions were employed to simulate lattice QCD at high temperature \cite{Cossu:2013uua,Chiu:2013wwa,Tomiya:2016jwr,Bazavov:2012qja,Buchoff:2013nra,Dick:2015twa,Mazur:2018pjw} (in Refs.~\cite{Dick:2015twa,Mazur:2018pjw}, only for valence quark sector).
JLQCD Collaboration studied with $N_f=2$ chiral fermions~\cite{Cossu:2013uua,Tomiya:2016jwr}.
In Ref.~\cite{Cossu:2013uua}, we generated the gauge ensembles with dynamical overlap fermions and applied a topology fixed approach at the $Q=0$ sector.
In Ref.~\cite{Tomiya:2016jwr}, gauge ensembles are generated with the M\"obius domain-wall (MDW) fermions \cite{Brower:2005qw,Brower:2012vk}, and a overlap/domain-wall reweighting technique \cite{Fukaya:2013vka,Tomiya:2016jwr} was applied, where observables measured on MDW fermion ensembles are reweighted to those on overlap fermion ensembles.
A disappearance of the $U(1)_A$ anomaly (at around $1.2T_c$) was also reported in simulations with $N_f=2$ non-chiral fermions by other groups \cite{Brandt:2016daq,Ishikawa:2017nwl}.
In Ref.~\cite{Bazavov:2019www}, they found that the $U(1)_A$ symmetry is good at $1.3 T_c$ but not near $T_c$.

\begin{table}[b!]
\vspace{-10pt}
  \centering
\caption{Numerical parameters of lattice simulations.
$L^3 \times L_t $ and $m$ are the lattice size and quark mass, respectively.
$\bar{\Delta}_{\pi-\delta}^{\mathrm{ov}}$ and $\chi_t$ are our results of the $U(1)_A$ susceptibility and topological susceptibility from the fermionic definition, respectively.
}
\small
\begin{tabular}{cccc}
\hline\hline
$L^3 \times L_t $ &  $am$  & $\bar{\Delta}_{\pi-\delta}^{\mathrm{ov}} a^2$ on OV & $\chi_t a^4$ \\
\hline
$24^3 \times 12$ & 0.001   & 1.5(0.6) $\times 10^{-6}$  & $\approx 0$ \\
$24^3 \times 12$ & 0.0025  & 3.6(1.3) $\times 10^{-5}$ & 5.0(3.7) $\times 10^{-8}$ \\
$24^3 \times 12$ & 0.00375 & 0.00017(7) & 2.3(0.7) $\times 10^{-7}$ \\
$24^3 \times 12$ & 0.005   & 0.00091(42) & 9.0(2.0) $\times 10^{-7}$ \\
$24^3 \times 12$ & 0.01    & 0.00389(92)& 1.7(0.2) $\times 10^{-6}$ \\
\hline
$32^3 \times 12$ & 0.001   & 1.8(1.4) $\times 10^{-5}$ & 8.8 (8.8) $\times 10^{-12}$ \\
$32^3 \times 12$ & 0.0025  & 0.00017(6)  & 3.5(3.0) $\times 10^{-8}$ \\
$32^3 \times 12$ & 0.00375 & 0.00026(8)  & 7.9(3.0) $\times 10^{-8}$ \\
$32^3 \times 12$ & 0.005   & 0.00291(188)& 9.3(1.9) $\times 10^{-7}$  \\
$32^3 \times 12$ & 0.01    & 0.01358(263)& 2.9(0.4) $\times 10^{-6}$  \\
\hline
$40^3 \times 12$ & 0.005   & 0.00785(178)& 5.4(0.6) $\times 10^{-7}$ \\
$40^3 \times 12$ & 0.01    & 0.01162(140)& 2.0(0.2) $\times 10^{-6}$ \\
\hline
$48^3 \times 12$ & 0.001   & 2.2(0.9)$\times 10^{-6}$ & 4.2(4.3) $\times 10^{-16}$ \\
$48^3 \times 12$ & 0.0025  & 0.00012(4)  & 4.9(4.4) $\times 10^{-9}$ \\
$48^3 \times 12$ & 0.00375 & 0.00032(12)  & 1.5(0.7) $\times 10^{-7}$ \\
$48^3 \times 12$ & 0.005   & 0.00135(63) & 2.9(1.1) $\times 10^{-7}$ \\
\hline\hline
\end{tabular}
\vspace{-10pt}
\label{Tab:param}
\end{table}

In these proceedings, we report on our recent results of the observables at $T=220 \ \mathrm{MeV}$ such as the Dirac spectrum, $U(1)_A$ susceptibility, screening masses from mesonic correlators, and topological susceptibility in $N_f=2$ lattice QCD simulations.
The simulation parameters are summarized in Table \ref{Tab:param}.
Our gauge ensembles are generated with the tree-level Symanzik improved gauge action and dynamical MDW fermions.
We use the gauge coupling $\beta=4.30$ and the lattice spacing $1/a=2.64 \ \mathrm{GeV}$ ($a \sim 0.075 \ \mathrm{fm}$), which is finer than that of configurations used in the previous works~\cite{Cossu:2013uua,Tomiya:2016jwr}.
We simulate lattice volumes $L=24,32,40,48$, and the length of the fifth dimension in the MDW fermion formulation is $L_s=16$.
The physical quark mass (as the average of up and down quark masses) is estimated to be $am=0.0014(2)$ ($3.7(5) \ \mathrm{MeV}$).
Some of our results were already reported in previous proceedings~\cite{Aoki:2017xux,Suzuki:2017ifu,Suzuki:2018vbe,Suzuki:2019vzy}.

\begin{figure}[t!]
    \vspace{-20pt}
    \centering
    \begin{minipage}[t]{0.5\columnwidth}  
            \includegraphics[clip,width=1.0\columnwidth]{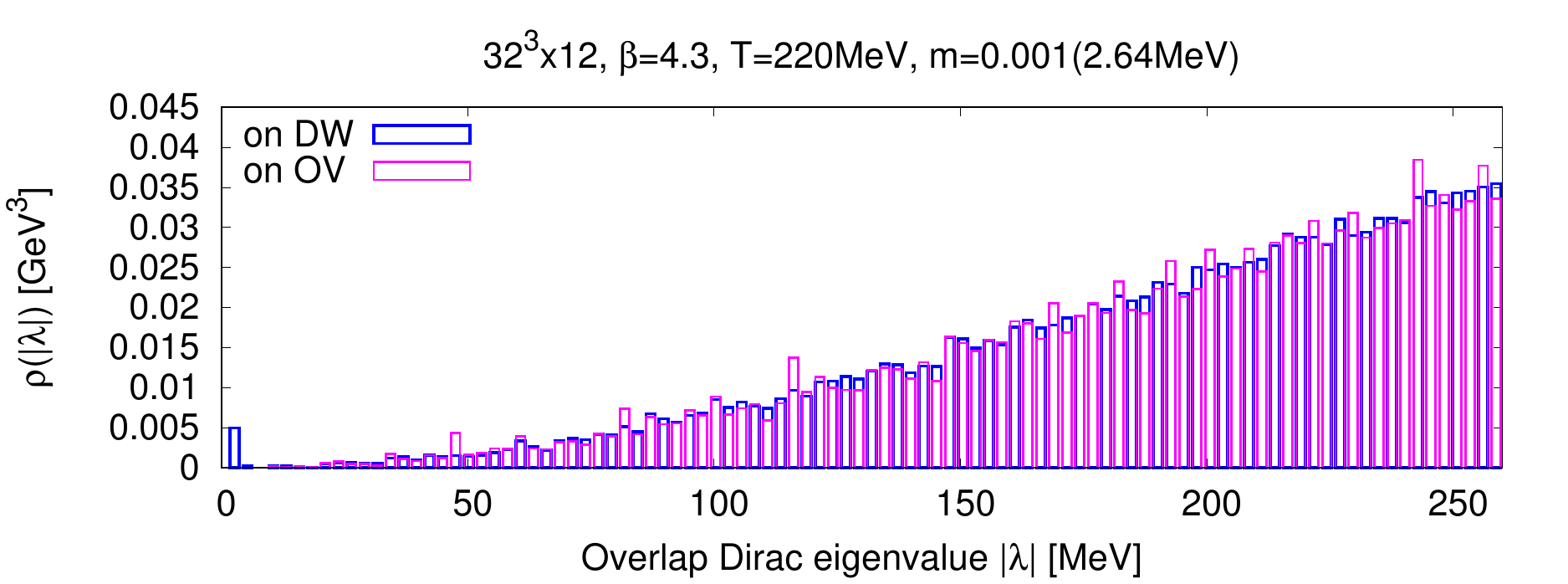}
    \end{minipage}
    \begin{minipage}[t]{0.5\columnwidth}
            \includegraphics[clip,width=1.0\columnwidth]{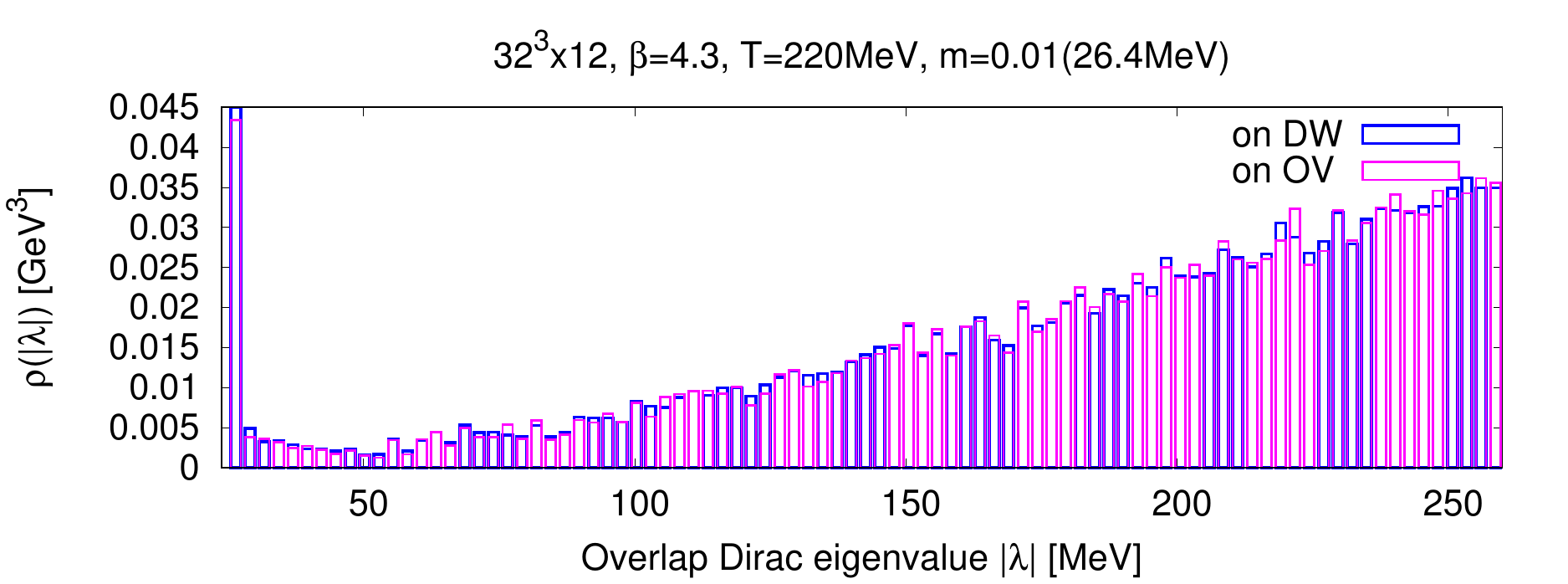}
    \end{minipage}
    \vspace{-5pt}
    \caption{Spectral density $\rho(|\lambda|)$ for overlap-Dirac eigenvalues $\lambda$ at $T=220 \ \mathrm{MeV}$.
Upper panel: $m=2.64 \ \mathrm{MeV}$.
Lower panel: $m=26.4 \ \mathrm{MeV}$.
 }
\vspace{-5pt}
    \label{fig:spe}
\end{figure}

\vspace{-5pt}
\section{Overlap Dirac spectrum}\label{sec-2}
\vspace{-5pt}
In Fig.~\ref{fig:spe}, we plot spectral density of overlap Dirac eigenvalues,  $\rho(\lambda)=(1/V)\langle\sum_{\lambda'}\delta(\lambda-\lambda')\rangle$ for two typical ensembles.
The blue and magenta bins denote the spectra on the MDW fermions ensembles (DW) and reweighted overlap fermion ensembles (OV), respectively.
At $m=2.64 \ \mathrm{MeV}$ for the OV ensembles, we find a suppression of both low eigenmodes and chiral zero modes.
The suppression of the low eigenmodes is related to the $U(1)_A$ symmetry restoration in the light quark mass region.
The disappearance of the chiral zero modes is related to the suppression of the topological susceptibility.
At $m=26.4 \ \mathrm{MeV}$, low eigenmodes are enhanced, which is related to the $U(1)_A$ symmetry breaking.

\vspace{-5pt}
\section{$U(1)_A$ susceptibility}\label{sec-3}
\vspace{-5pt}
The $U(1)_A$ susceptibility $\Delta_{\pi - \delta}$ is an order parameter of the $U(1)_A$ symmetry breaking.
This is defined from a spacetime integral of the difference between two-point correlators of isovector-pseudoscalar ($\pi^a \equiv i \bar{\psi} \tau^a \gamma_5 \psi$) and isovector-scalar ($\delta^a \equiv \bar{\psi} \tau^a \psi$) operators:
\begin{equation}
\Delta_{\pi-\delta} \equiv \chi_\pi - \chi_\delta \equiv \int d^4x \langle \pi^a(x) \pi^a(0) - \delta^a (x) \delta^a(0) \rangle, \label{eq:Delta_def}
\end{equation}
where $a$ is an isospin index in $N_f=2$ QCD.
The $U(1)_A$ susceptibility in the lattice theory is defined by a summation of low-lying eigenvalues of the overlap Dirac operator, $\lambda_i^{(\mathrm{ov},m)}$ \cite{Cossu:2015kfa}:
\begin{equation}
\Delta_{\pi-\delta}^{\mathrm{ov}} =  \frac{1}{V(1-m^2)^2} \left< \sum_i \frac{2m^2(1-\lambda_i^{(\mathrm{ov},m)2})^2}{\lambda_i^{(\mathrm{ov},m)4}} \right> , \label{eq:Delta_ov}
\end{equation}
where we set the lattice spacing $a=1$.
This summation is truncated at the lowest 40 eigenvalues.\footnote{From this definition, we further apply two types of subtractions: a subtraction of the contributions from chiral zero modes and an ultraviolet divergence (or lattice cutoff).
For a justification of the zero mode subtraction, see Ref.~\cite{Aoki:2012yj,Tomiya:2016jwr}.
For the parametrization scheme of the lattice cutoff contribution by different valence quark masses, see Ref.~\cite{Suzuki:2018vbe,Suzuki:2019vzy}.}

\begin{figure}[t!]
    \vspace{-10pt}
    \begin{center}
            \includegraphics[clip,width=0.5\columnwidth]{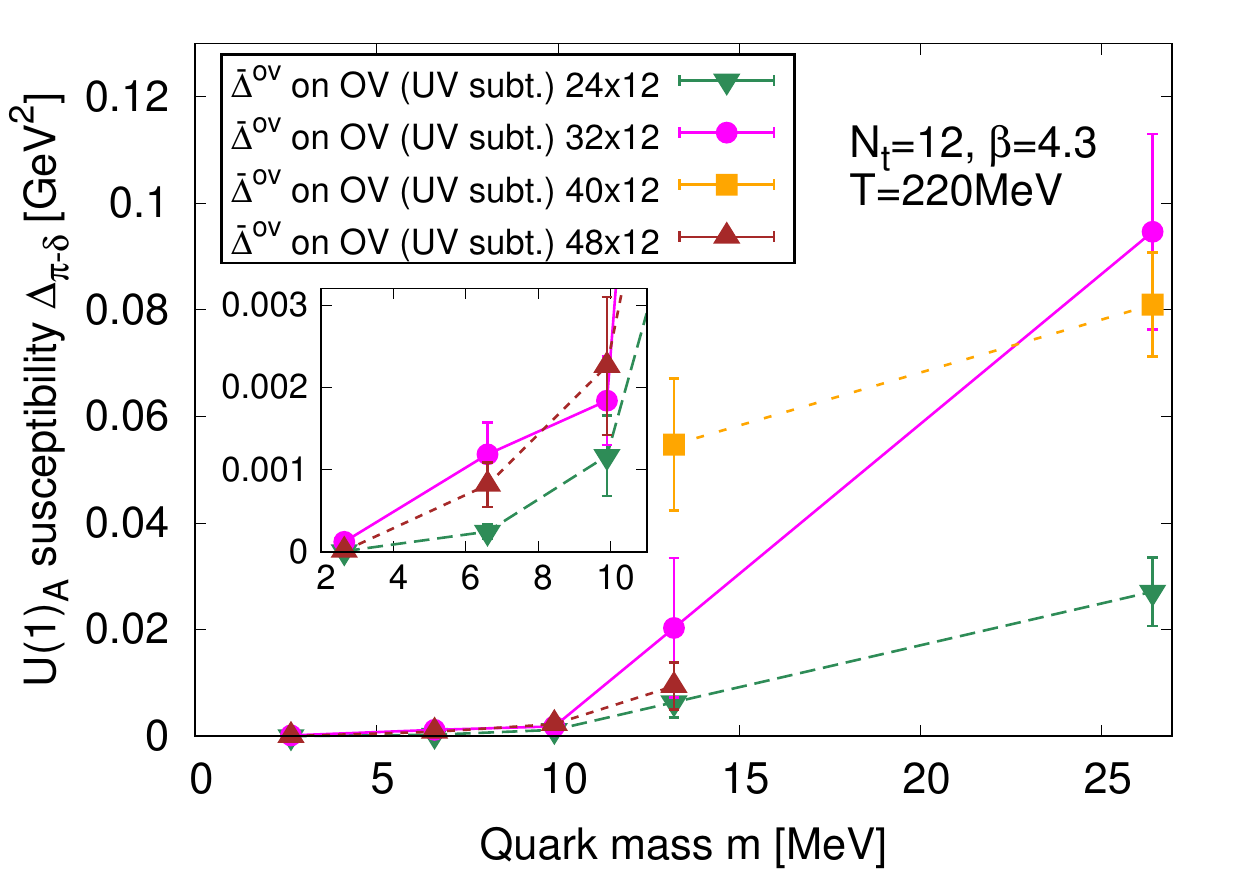}
    \end{center}
    \vspace{-15pt}
    \caption{$U(1)_A$ susceptibilities, $\bar{\Delta}_{\pi-\delta}^\mathrm{ov}$ (\ref{eq:Delta_ov}), from the eigenvalue density of the overlap-Dirac operators at $T=220 \ \mathrm{MeV}$.}
    \label{fig:U1A}
    \vspace{-10pt}
\end{figure}

In Fig.~\ref{fig:U1A}, we show the $U(1)_A$ susceptibility at $T=220 \ \mathrm{MeV}$.
In the light quark mass region, we find strong suppression of the $\Delta_{\pi-\delta}^{\mathrm{ov}}$.
For example, at the lowest quark mass and $L=32$, the ratio of $\Delta_{\pi-\delta}^{\mathrm{ov}}$ to temperature is $\sqrt{\Delta_{\pi-\delta}^{\mathrm{ov}}}/T \approx 5\%$.
The volume dependence is small for $L=$ 24--48.
The data at different volumes are consistent except for the heaviest quark mass at $L=24$, whose aspect ratio against temperature is $L/L_t=2$.

\section{Screening mass difference from spatial mesonic correlators}\label{sec-4}

The screening mass is defined by the exponential decay of spatial correlators", which may be used to measure a violation of $U(1)_A$ symmetry.
We investigate the difference between the effective screening masses
\begin{equation}
 \Delta m_{scr} (z) = | m_{scr}^{PS}(z) - m_{scr}^{S}(z)|, \label{eq:m_scr}
\end{equation}
where $m_{scr}^{PS} (z)$ and $m_{scr}^{S} (z)$ are the effective screening masses at a spatial coordinate $z$ for isovector-pseudoscalar ($\pi^a \equiv i \bar{\psi} \tau^a \gamma_5 \psi$) and isovector-scalar ($\delta^a \equiv \bar{\psi} \tau^a \psi$) operators, respectively.

\begin{figure}[t!]
    \vspace{-10pt}
    \begin{center}
           \includegraphics[clip,width=0.5\columnwidth]{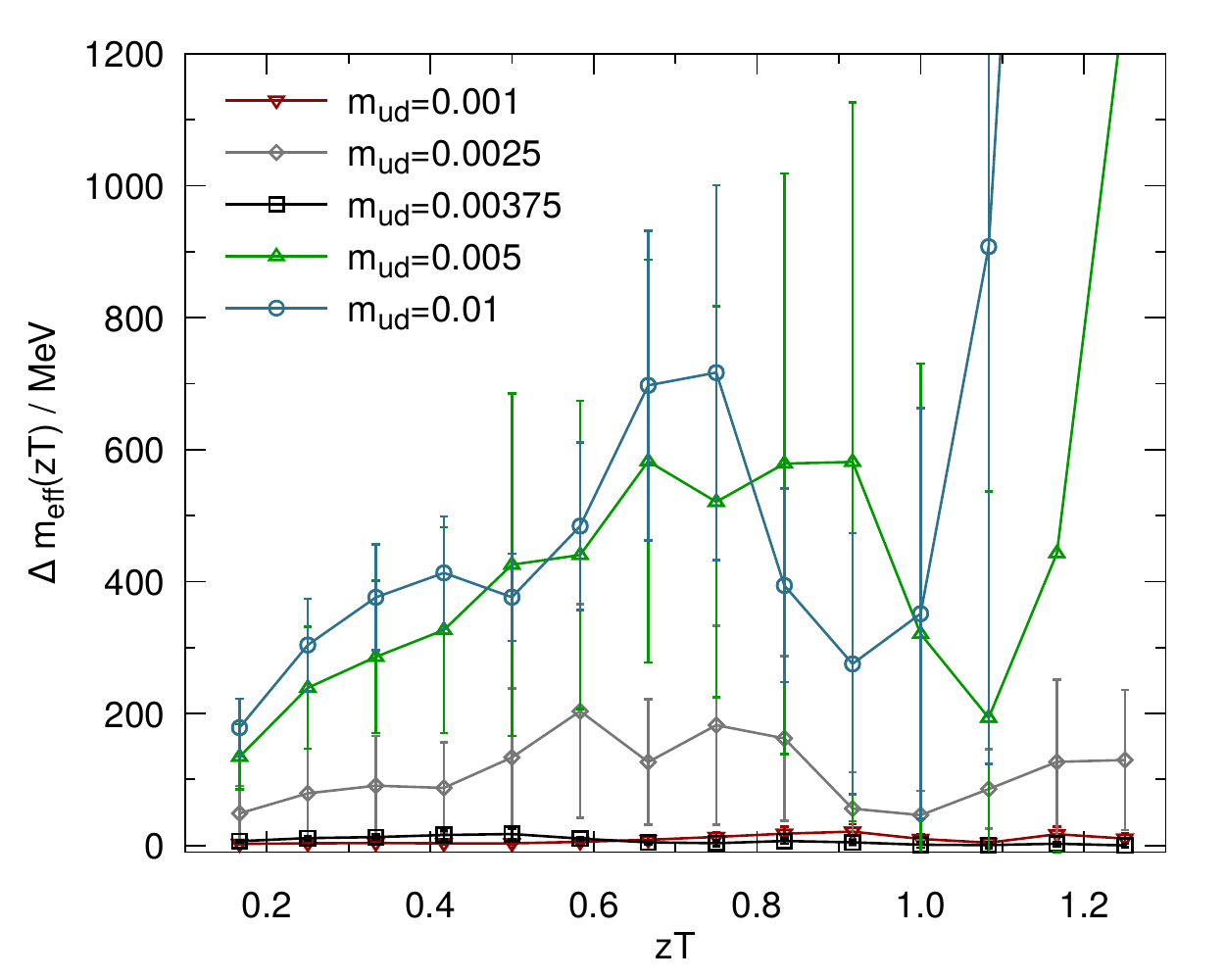}
    \end{center}
    \vspace{-15pt}
    \caption{Difference between effective screening masses (\ref{eq:m_scr}) from spatial mesonic collerators for $U(1)_A$ partners at $T=220 \ \mathrm{MeV}$ and $L=32$.
The horizontal axis is defined as a dimensionless spatial distance $zT=(n_za/N_ta) =n_z/N_t$.}
    \label{fig:massdiff}
    \vspace{-10pt}
\end{figure}

In Fig.~\ref{fig:massdiff}, we show the difference between the effective screening masses measured by the MDW operator (without reweighting), where the horizontal axis is a dimensionless spatial distance ($zT=(n_za/N_ta) =n_z/N_t$).
For the screening masses with light quark mass, we find a small value of $\Delta m_{scr} (zT)$, which indicate the restoration of the $U(1)_A$ symmetry and it is consistent with the results of the $U(1)_A$ susceptibility $\bar{\Delta}_{\pi-\delta}^\mathrm{ov}$.
For heavy quark masses, the mass difference becomes large, which implies the $U(1)_A$ symmetry breaking.

\vspace{-5pt}
\section{Topological susceptibility}\label{sec-5}
\vspace{-5pt}

The topological susceptibility $\chi_t$ is defined as a gauge ensemble average of the topological charge $Q_t$:
\begin{equation}
\chi_t=\frac{\langle Q_t^2 \rangle}{V}, \label{chit}
\end{equation}
For the topological charge $Q_t$, we employ two definitions.
As a fermionic definition, $Q_t$ is defined through the index theorem for the overlap Dirac operator:
\begin{equation}
Q_t=n_+ - n_-,  \label{Q_fermion}
\end{equation} where $n_\pm$ are the numbers of chiral zero modes with positive or negative chirality, respectively.
As a gluonic definition, $Q_t$ is defined as a summation over spacetime $x$ at a flow time $t$ :
\begin{equation}
Q_t (t)=\frac{1}{32\pi^2} \sum_x \varepsilon^{\mu \nu \rho \sigma} \mathrm{Tr} \, F_{\mu \nu}(x,t) F_{\rho \sigma} (x,t), \label{Q_gluon}
\end{equation}
where $F_{\mu \nu} (x,t)$ is the clover-type discretization of the field strength tensor \cite{Bruno:2014ova}.\footnote{
This definition is usually not an integer, but we find a well-discretized distribution of $Q_t (t)$ at $t=5$.}

\begin{figure}[t!]
\vspace{-20pt}
    \begin{center}
            \includegraphics[clip,width=0.5\columnwidth]{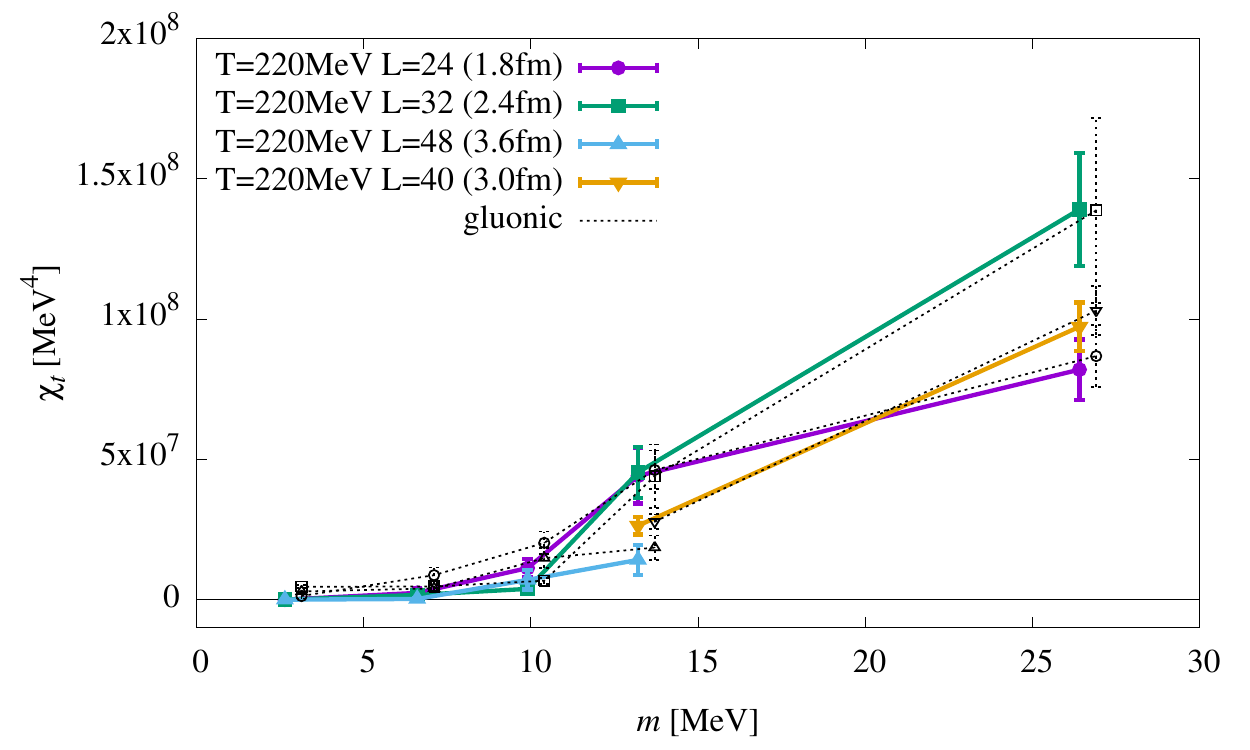}
    \end{center}
    \vspace{-20pt}
    \caption{Topological susceptibilities $\chi_t$ at $T=220 \ \mathrm{MeV}$.
    Colored points: $\chi_t$ from the fermionic definition (\ref{Q_fermion}) on reweighted OV ensembles.
    Uncolored points: $\chi_t$ from the gluonic defnition (\ref{Q_gluon}) on MDW ensembles.
}
    \label{fig:chit}
    \vspace{-10pt}
\end{figure}

In Fig.~\ref{fig:chit}, we plot the topological susceptibility $\chi_t$ at $T=220 \ \mathrm{MeV}$.
We show the results from the fermionic definition (\ref{Q_fermion}) on the OV ensembles and the gluonic definition (\ref{Q_gluon}) on the MDW ensembles, respectively.
In the light quark mass region, $\chi_t$ is strongly suppressed with both the definitions.
Furthermore, the volume dependence between $L=24$ and $48$ is small.
In the heavy quark mass region, the value of $\chi_t$ becomes nonzero, which is in agreement with the peak structure of the Dirac spectra in the lower panel of Fig.~\ref{fig:spe}.

\vspace{-10pt}
\section{Summary and discussion}\label{sec-6}
\vspace{-5pt}

In these proceedings, we studied the high-temperature phase of QCD at $T=220 \ \mathrm{MeV}$ by using $N_f=2$ lattice QCD simulations with dynamical MDW fermions.
We found small values of the $U(1)_A$ susceptibility (\ref{eq:Delta_ov}) and the difference of mesonic screening masses (\ref{eq:m_scr}) in light quark mass region, $m \lesssim 10 \ \mathrm{MeV}$, which indicates the $U(1)_A$ symmetry restoration in the chiral limit ($m \to 0$).
Furthermore, we found strong suppression of the topological susceptibility in the light-quark mass region.
The mesonic and baryonic correlators at higher temperature were already reported in Refs.~\cite{Rohrhofer:2017grg,Rohrhofer:2019qwq,Rohrhofer:2019yko}.

\vspace{-10pt}
\section*{Acknowledgment}\label{sec-Ack}
\vspace{-10pt}
Numerical simulations are performed on IBM System Blue Gene Solution at KEK under a support of its Large Scale Simulation Program (No. 16/17-14) and Oakforest-PACS at JCAHPC under a support of the HPCI System Research Projects (Project IDs: hp170061, hp180061 and hp190090) and Multidisciplinary Cooperative Research Program in CCS, University of Tsukuba (Project IDs: xg17i032 and xg18i023).
This work is supported in part by the Japanese Grant-in-Aid for Scientific Research (No. JP26247043, JP18H01216 and JP18H04484), and by MEXT as ``Priority Issue on Post-K computer" (Elucidation of the Fundamental Laws and Evolution of the Universe) and by Joint Institute for Computational Fundamental Science (JICFuS).

\vspace{-10pt}
\bibliographystyle{JHEP}
\bibliography{lattice2019}
\vspace{-15pt}
\end{document}